\documentclass[12pt]{iopart}

\usepackage{iopams}  
\usepackage{graphicx}  
\begin{document}


\title[First results from Au+Au collisions at $\sqrt{s_{NN}}$ = 9.2 GeV in STAR]
{First results from Au+Au 
collisions at $\sqrt{s_{NN}}$ = 9.2 GeV in STAR}

\author{Lokesh Kumar (for the STAR Collaboration)}

\address{Department of Physics, \\
Panjab University, \\ 
Chandigarh, \\
INDIA - 160014.}
\ead{lokesh@rcf.rhic.bnl.gov}
\begin{abstract}
One of the primary aims of heavy-ion collisions is to map the
QCD phase diagram and search for different phases and phase boundaries. 
RHIC Energy Scan Program was launched to address this goal by studying 
heavy-ion collisions at different center of mass energies. The first 
test run with Au+Au collisions  at $\sqrt{s_{NN}}$ = 9.2 GeV took 
place in early 2008. The large acceptance STAR detector has collected 
few thousands minimum bias collisions at this beam energy.
We present the first results for identified particle 
yields and particle ratios. The results for the azimuthal 
anisotropy parameters $v_1$ and $v_2$ and those from pion interferometry
measurements are also discussed in this paper. These results 
are compared to data from the SPS at similar beam energies.

\end{abstract}


\section{Introduction: }
One of the main aims of heavy-ion collision experiments is to map the Quantum
Chromodynamics (QCD) phase diagram~\cite{star_white_paper}. The QCD phase diagram is described 
by the variation of temperature with respect to the baryon chemical potential ($\mu_{B}$).
Efforts are being made to locate the QCD phase boundary (separates the matter 
dominated by quarks and gluons degrees of freedom from those dominated by hadronic degrees 
of freedom) and the QCD critical
point (where the first order phase transition ends) in the QCD phase diagram.
To locate these or to map the
QCD phase diagram, one needs to find a way to vary temperature and $\mu_{B}$, which
can be achieved by varying the colliding beam energy.
The freeze out temperature and $\mu_{B}$ can be
deduced from the spectra and
ratios of produced particles by comparing with model calculations.
Various signatures for locating
different phases of matter and the QCD critical point will be studied.

Relativistic Heavy Ion Collider (RHIC) Energy Scan Program will be launched to address
these issues and specifically, the STAR experiment (Solenoidal Tracker At RHIC) would also like to
determine the beam energy at which the following interesting phenomena observed at top
RHIC energy (200 GeV) start to appear in the data :
Number of constituent quark (NCQ) scaling of elliptic flow parameter for produced hadrons at
intermediate $p_{t}$,
enhanced correlated yields at large $\Delta\eta$ and $\Delta\phi$ $\sim$ 0 (Ridge) and
the suppression of high transverse momentum hadron production in heavy ion collisions
~\cite{qscale,ridge,suppression}.
To achieve these goals, STAR has proposed a beam energy scan program at RHIC spanning
beam energies from 5 GeV to 50 GeV. As a first step of the energy scan program, a test run
was organized in early 2008 with Au+Au collisions at $\sqrt{s_{NN}}$ = 9.2 GeV.

\section{Experiment and Analysis}
The data presented here are from Au+Au collisions at $\sqrt{s_{NN}}$ = 9.2 GeV using the
Time Projection Chamber (TPC)~\cite{star_nim} in the STAR experiment at RHIC.
The events with a primary vertex within $\pm$ 75 cm of the geometric center of TPC along
the beam axis were accepted for this analysis and about $\sim$3000 events were analyzed. 
The centrality selection for Au+Au 9.2 GeV was done by using the uncorrected
charged particles, measured event-by-event within the pseudo-rapidity ($\eta$) 
region $\pm$ 0.5. 
Those primary tracks which originate within 3 cm of the 
primary vertex (distance of closest approach or DCA) and have transverse momentum ($p_{t}$) $>$ 0.1 
GeV/c
within the rapidity (y) interval $|$y$|$ $<$ 0.5 were selected for 
the analysis. 

Particle identification is accomplished by measuring the ionization energy loss (dE$\slash$dx).
For good momentum and $dE$\slash$dx$
resolution, tracks were required to have at least
20 number of fit points (Nfitpts) out of the maximum 45 hits in the TPC. Uncorrected particle 
yields are extracted from 
$dE$\slash$dx$ distributions for each $p_{t}$, y and centrality 
bin~\cite{dedx}.

The spectra were corrected for tracking inefficiency, detector acceptance and energy loss
due to interactions. Total reconstruction efficiencies were obtained from embedding 
Monte-Carlo (MC) tracks into real events at the raw data level and subsequently reconstructing these
events. Detailed description of STAR geometry and a realistic simulation of TPC are given
in references~\cite{dedx}. 
The 0-10$\%$ centrality results presented here are not corrected for vertex inefficiency.
Simulations suggest that this effect is small for collision centrality studied.

The systematic uncertainties on yield of particles were obtained by varying the 
analysis cuts like 
vertex-z position, Nfitpts, DCA and y. The systematic errors due to the type of function used 
to fit the spectra (error due to extrapolation), error on efficiency and error in extracting 
the yields of particles were also taken into account. The total systematic error for pions 
were found to be $\sim$10\%, $\sim$12\% for kaons and greater than 15\% for protons (results not 
shown in this paper).

\section{Results} 
\begin{figure}
\begin{center}
\includegraphics[height=12pc,width=16pc]{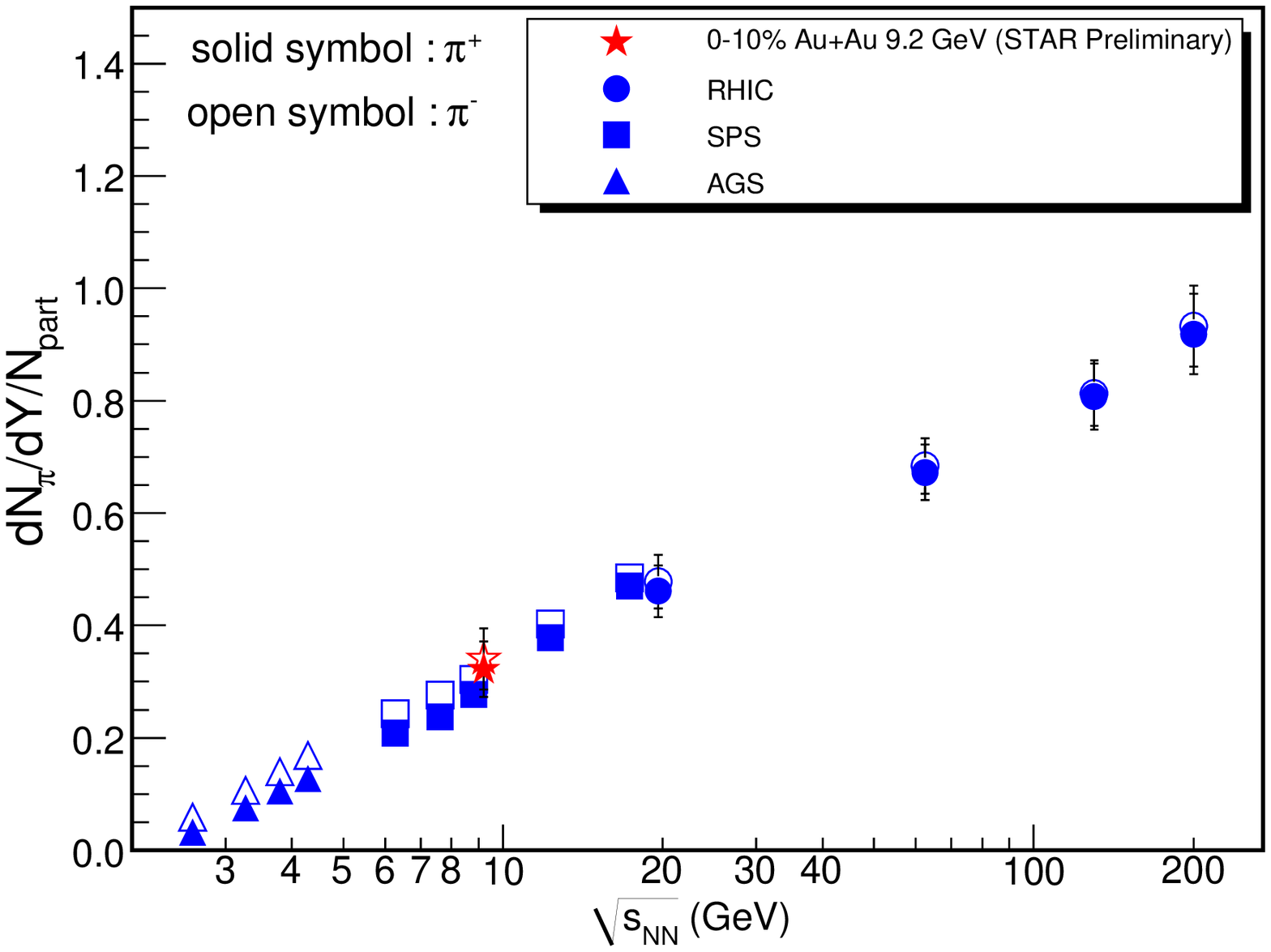}
\includegraphics[height=12pc,width=16pc]{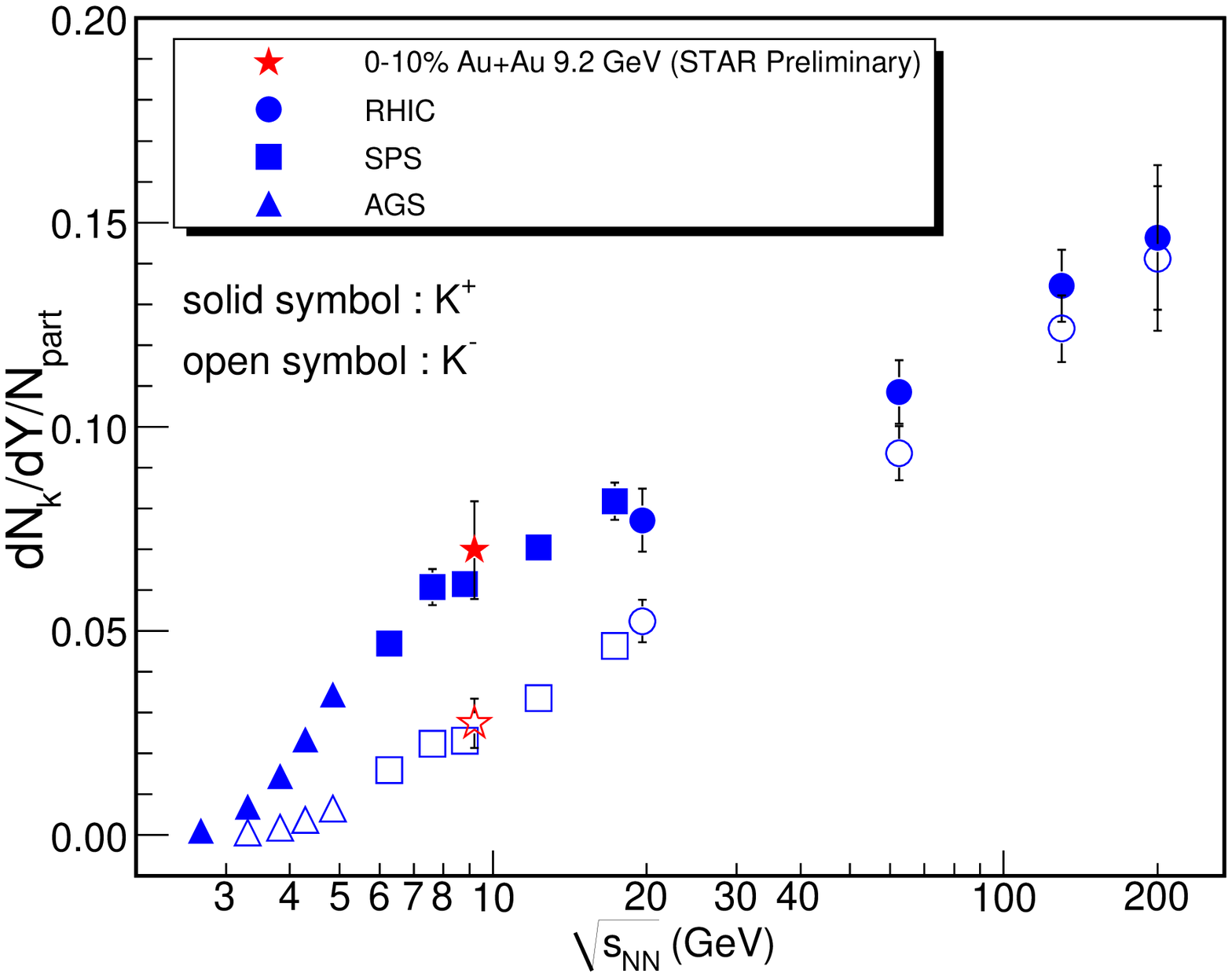}
\caption{\label{fig:epsart}
Left panel : $\pi^{\pm}$ yield per unit rapidity (dN/dy) as function of $\sqrt{s_{NN}}$  at 
mid-rapidity for
central collisions. The errors are statistical and systematic added in quadrature.
Right panel :  Same as above, for $K^{\pm}$.
Star symbols are the  results for Au+Au 9.2 GeV, circles represent 
RHIC results from other energies, squares show the results from SPS and triangles represent 
those from AGS energies (This is same for all figures).
RHIC : Results for $\sqrt{s_{NN}}$ = 9.2 GeV and 19.6 - 200 GeV; SPS : Results for $\sqrt{s_{NN}}$ 
= 6.3 - 17.3 GeV;
AGS : Results for $\sqrt{s_{NN}}$ $<$ 6 GeV.
}
\end{center}
\end{figure}

\subsection{Hadron Yields,  Average Transverse Mass and Ratios}
Fig.~\ref{fig:epsart} shows the particle yields at mid-rapidity dN/dy for $\pi^{\pm}$ and $K^{\pm}$
 plotted as a function of $\sqrt{s_{NN}}$ for central collisions at AGS, SPS and RHIC energies (as
mentioned in the caption)~\cite{sps_rhic}.
\begin{figure}
\begin{center}
\includegraphics[height=12pc,width=16pc]{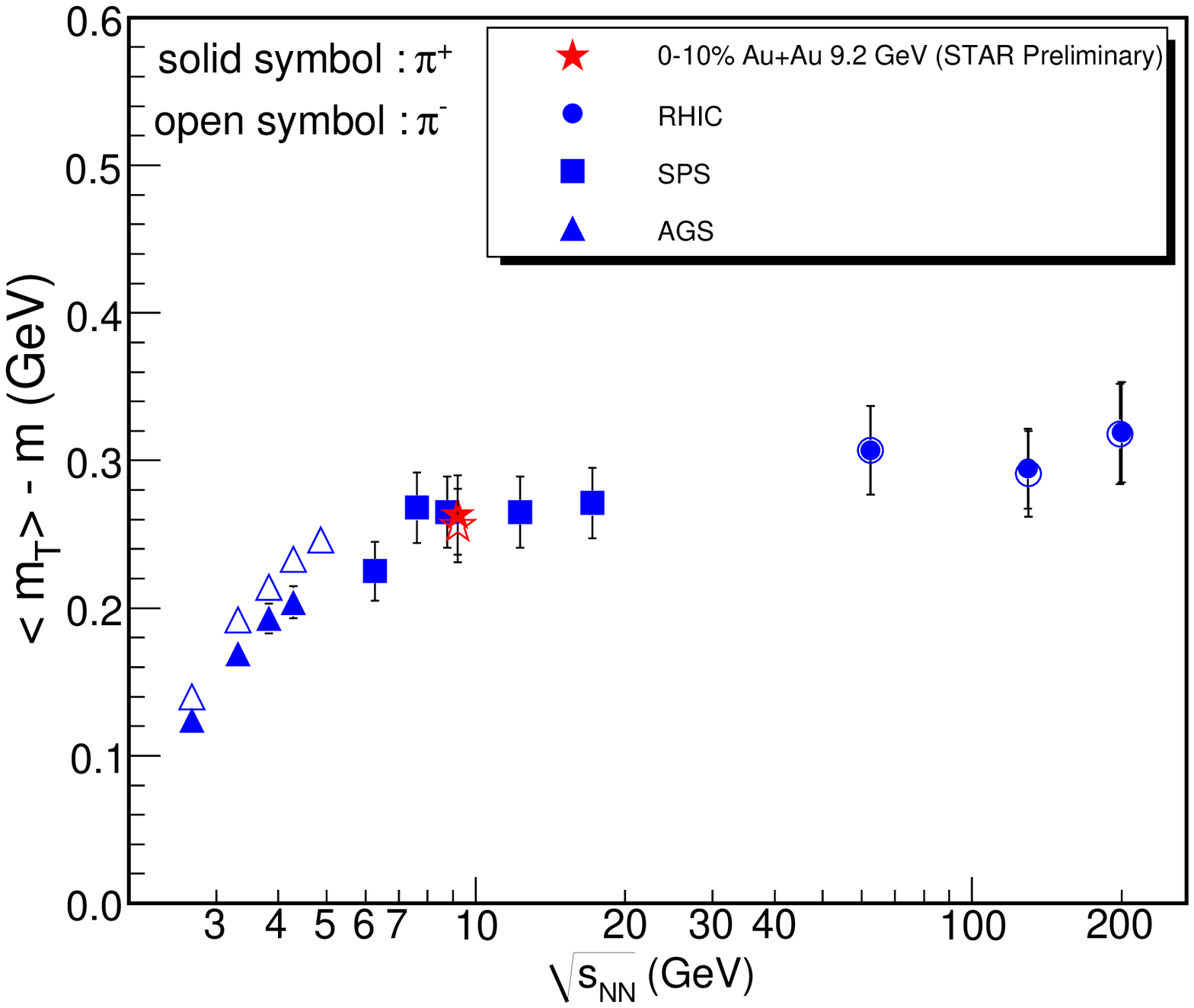}
\includegraphics[height=12pc,width=16pc]{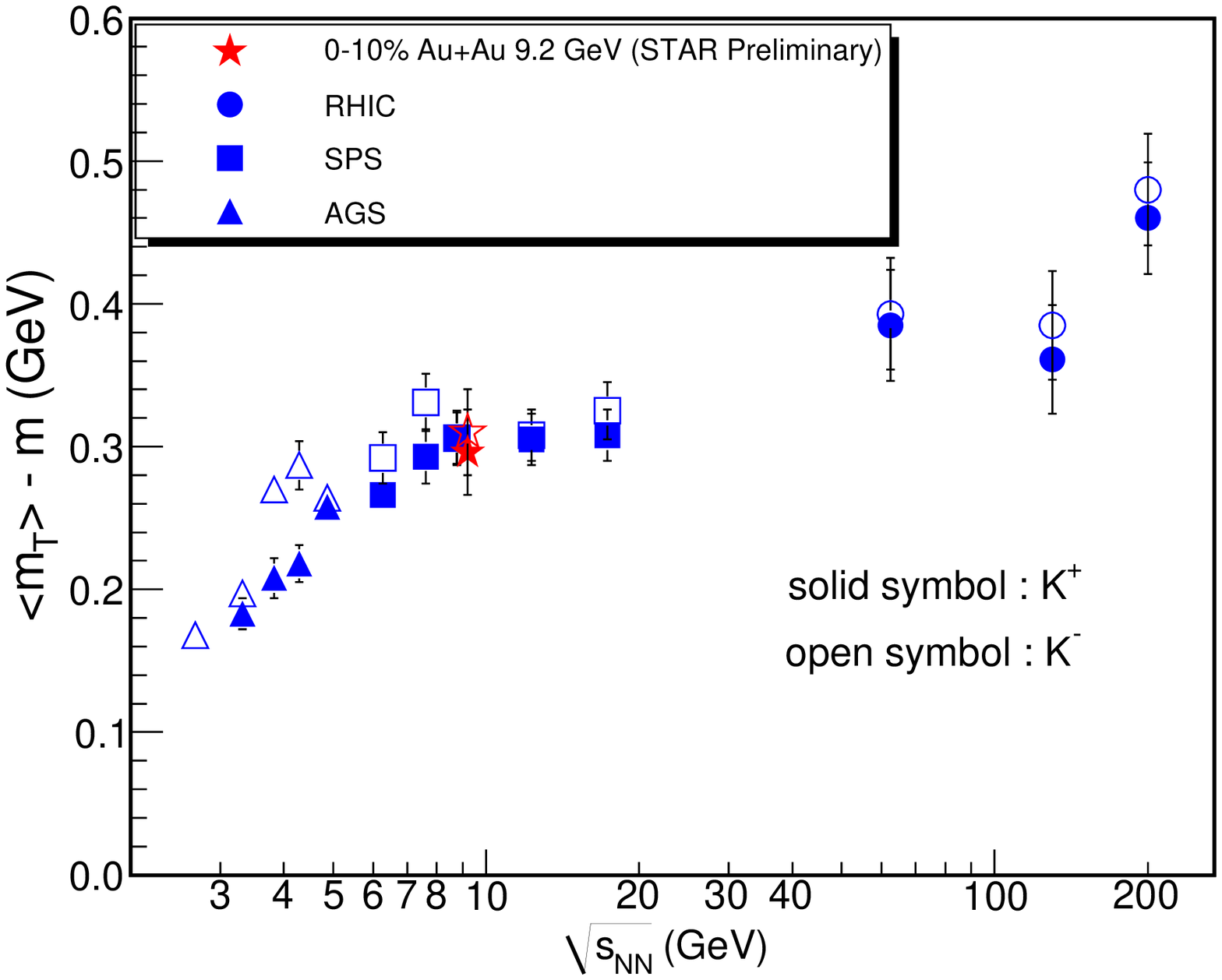}
\caption{\label{fig:epsart1} Left panel : Average transverse mass for $\pi^{\pm}$ as function of
colliding beam energy. Solid symbol are for the positively charged particles and open symbols
for the negatively charged particles. 
Errors are statistical and systematic added in quadrature.
Right panel : same as above, for $K^{\pm}$.
}
\end{center}
\end{figure}

Fig.~\ref{fig:epsart1} shows the difference between average transverse mass 
($<m_{T}>$) and hadron mass (m) plotted as a 
function of colliding beam energy ($\sqrt{s_{\mathrm {NN}}}$)~\cite{sps_rhic}. 
For a thermodynamical system, $<m_{T}>$ - m
is related to temperature of the system and $\log({\sqrt{s_{\mathrm {NN}}}})$,
which is proportional to yield of the particles, is related to entropy of
the system.

\begin{figure}
\begin{center}
\includegraphics[height=12pc,width=17.5pc]{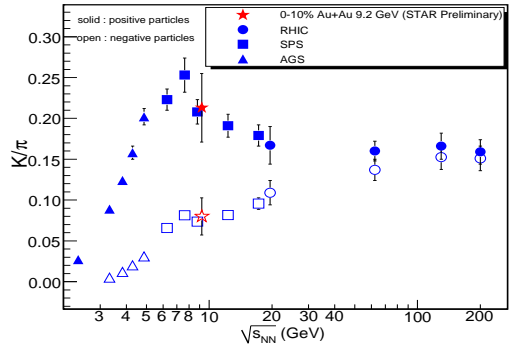}
\caption{\label{fig:epsart2} Charged kaons to charged pions ratio as a function of
colliding beam energy. Solid symbols represent $K^{+}\slash\pi^{+}$ ratio and the open 
symbols represent the $K^{-}\slash\pi^{-}$ ratio. 
Errors are statistical and systematic added in quadrature.
}
\end{center}
\end{figure}

The strangeness production in heavy-ion collision experiments can be observed from the 
kaon to pion ratio. Fig.~\ref{fig:epsart2}  shows the $K^{\pm}\slash\pi^{\pm}$ ratio as a function 
of $\sqrt{s_{\mathrm {NN}}}$. 

\subsection{Azimuthal Anisotropy Measurements}
\begin{figure}
\begin{center}
\includegraphics[scale=0.4]{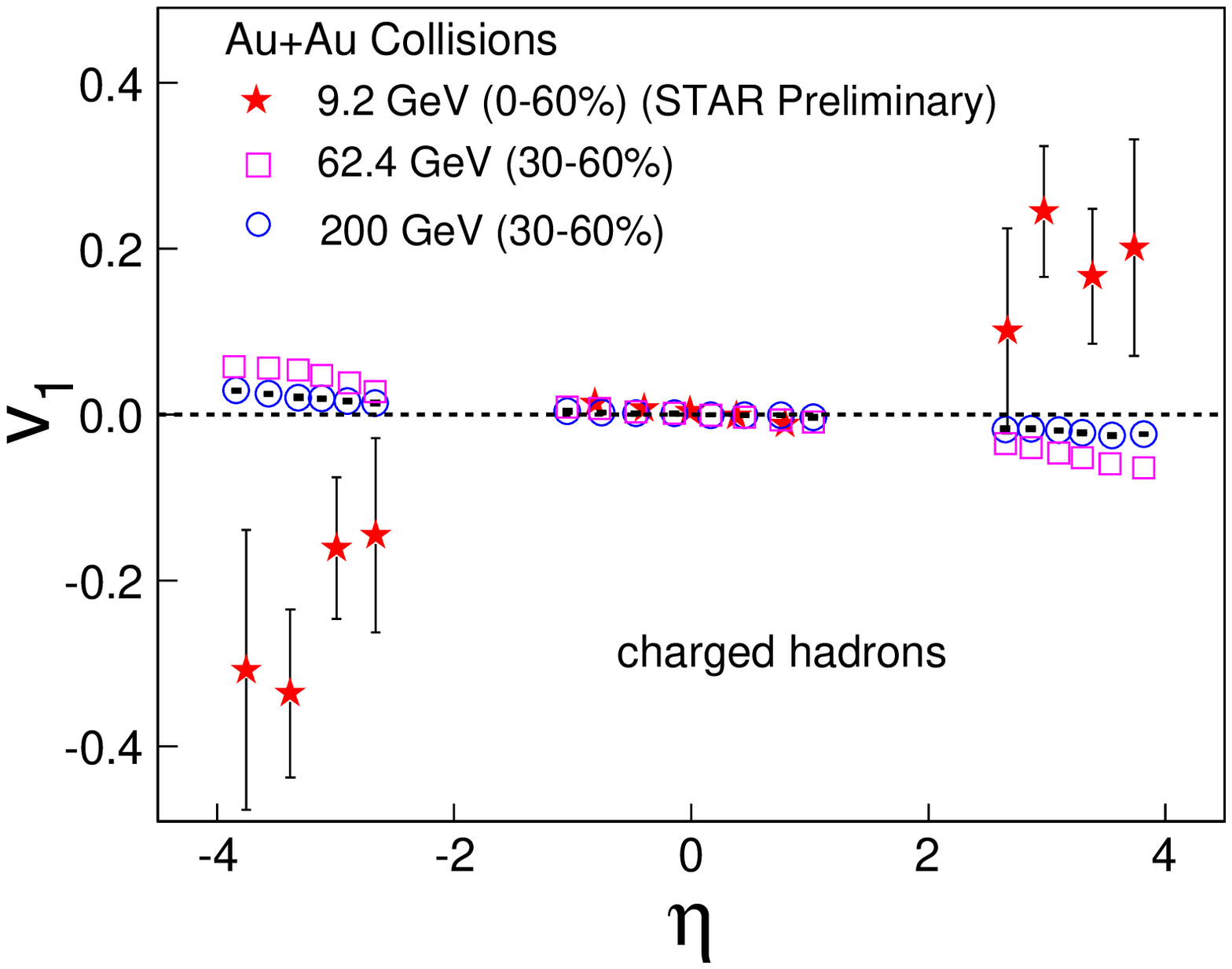}
\includegraphics[scale=0.3]{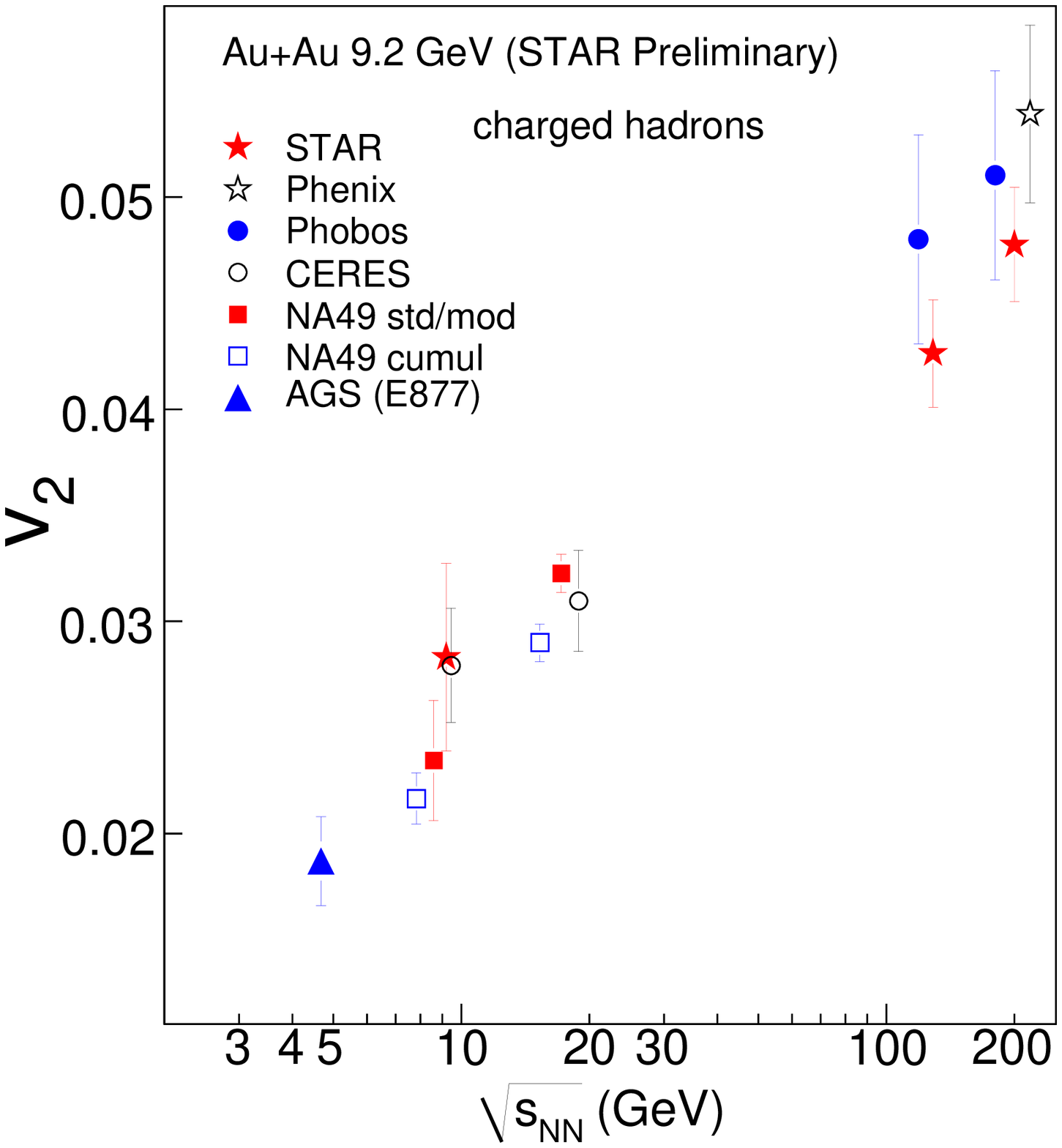}
\vspace{-0.5cm}
\caption{\label{fig:epsart3} Left panel : $v_{1}$ as function of $\eta$ for charged hadrons.
Results for Au+Au collisions at $\sqrt{s_{NN}}$ = 9.2 GeV are compared with those from 200
and 62.4 GeV.
Right panel : $v_{2}$ (0-60\% for 9.2 GeV data) for charged hadrons as function of colliding beam energy.
Errors are statistical only.
}
\end{center}
\end{figure}

Fig.~\ref{fig:epsart3} (left panel) shows the directed flow ($v_{1}$) as a function of
$\eta$ for charged hadrons. Results for Au+Au collisions at $\sqrt{s_{NN}}$ = 9.2 GeV 
are compared with those from 200 and 62.4 GeV~\cite{direct}. It is observed that $v_{1}$ for 9.2 GeV
shows different behavior as compared to 200 and 62.4 GeV. This could be due to the
spectators effect, since the beam rapidity for 9.2 GeV is 2.3, which is within the
Forward Time Projection Chamber (FTPC) acceptance in the STAR experiment. For 200 and 62.4 GeV,
the beam rapidities are 5.4 and 4.2 respectively, which lie outside the FTPC acceptance region and hence
flow due to spectators is not observed. Fig.~\ref{fig:epsart3} (right panel) shows the
$v_{2}$ as a function of beam energy for charged hadrons~\cite{qscale,ellipt}.

\subsection{Pion Interferometry Measurements}
\begin{figure}
\begin{center}
\includegraphics[scale=0.3]{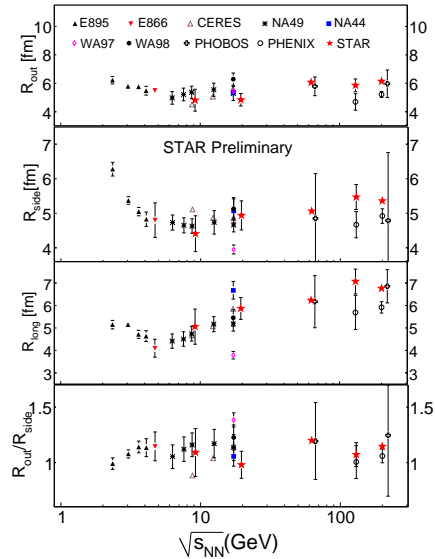}
\vspace{-0.4cm}
\caption{\label{fig:epsart4} : The HBT radii $R_{out}$, $R_{side}$, $R_{long}$ and the ratio
$R_{out}/R_{side}$ for $\pi^{+}$ are plotted as function of beam energy.
The errors for Au+Au 9.2 GeV are only statistical. The systematic errors are expected 
to be less than $\sim$10\% for each out, side and long.
}
\end{center}
\end{figure}

Fig.~\ref{fig:epsart4} shows the results for HBT (Hanbury Brown and Twiss) pion ($\pi^{+}$) 
interferometry measurements. In HBT 
measurement, we measure three radii - $R_{out}$, $R_{side}$ and $R_{long}$. $R_{out}$ 
measures the spatial and temporal extension of the source, $R_{side}$ measures the 
spatial extension of the source and the ratio $R_{out}/R_{side}$ gives the emission duration
of the source. It is expected that for a first order phase transition, the ratio $R_{out}/R_{side}$ is
very large compared to 1~\cite{hbt}. Fig.~\ref{fig:epsart4} shows that for 
the measured beam energies, the ratio $R_{out}/R_{side}$ is close to 1. 

\section{Conclusions and Summary}
We observe that the hadronic yields, $<m_{T}>$ - m and ratios
for Au+Au 9.2 GeV data presented are consistent with the observed beam energy dependence within
the errors, which are dominated by statistics collected in this test run at RHIC. 
The results for azimuthal anisotropy measurements ($v_{1}$, $v_{2}$) are obtained in the Au+Au collisions
at $\sqrt{s_{NN}}$ = 9.2 GeV. The $v_{2}$ results are similar to those obtained at SPS from 
collisions at similar
beam energies and follow the observed beam energy dependence. It is observed that results for
pion interferometry measurements for Au+Au 9.2 GeV also follow the established beam energy
dependence.
These results are from only $\sim$3000 good events and the higher statistics 
will help doing a significant qualitative improvement. Apart from this,
the Collider environment provides significant advantages over fixed target experiments. 
The $p_{t}$ vs. rapidity acceptance for produced particles 
are uniform over all beam energies. Particle density per unit area are
much reasonable and slowly varying with beam energy. In STAR there is added
advantage of excellent particle identification, which will improve further
with addition of Time-Of-Flight system in the year 2010.


In summary, we have obtained the results for identified hadron spectra, azimuthal
anisotropy and pion interferometry measurements for Au+Au collisions 
at 9.2 GeV, the lowest beam energy injected at RHIC so far. 
The results from this small statistics data indicate that various observables 
follow the previously established beam energy dependence. 
Through this test run we have shown 
the readiness of STAR experiment for the 
future Beam Energy Scan program at RHIC. The higher
statistics and good particle identification capability in the STAR experiment 
in a Collider set-up will help in locating the critical point and 
map the QCD phase diagram along with
some new measurements that were not possible in SPS. 
It will also help locate the $\sqrt{s_{NN}}$ where the
onset of several interesting observations (NCQ scaling of $v_{2}$~\cite{qscale},
high $p_{t}$ hadron suppression in A+A collisions relative to p+p
collisions~\cite{suppression} and the ridge formation~\cite{ridge}) in Au+Au collisions at
 $\sqrt{s_{NN}}$ = 200 GeV at RHIC.

\end{document}